\documentclass{amsart}
\usepackage{amsfonts}
\usepackage[utf8x]{inputenc}
\usepackage{graphicx,epsfig}
\usepackage{amsmath,amsbsy}
\usepackage{amsfonts}
\usepackage{amssymb}
\usepackage[T1]{fontenc}
\usepackage{graphicx}
\usepackage{amscd}

\newtheorem{theorem}{Theorem}
\theoremstyle{plain}

\newtheorem{lemma}{Lemma}

\numberwithin{equation}{section}

\begin{document}

\title[Generalized Pickands]{On the Pickands stochastic process}

\author{Adja Mbarka Fall$^{*}$ and Gane Samb LO$^{**}$}
\address{$^{*}$ LERSTAD, Universit\'e Gaston Berger de Saint-Louis, SENEGAL%
gane-samb.lo@ugb.edu.sn, ganesamblo@ufrsat.org}
\address{$^{**}$ LSTA, UPMC, France and LERSTAD, Universit\'e Gaston Berger
de Saint-Louis, SENEGAL.}


\begin{abstract}
We consider the Pickands process
\begin{equation*}
P_{n}(s)=\log (1/s)^{-1}\log \frac{X_{n-k+1,n}-X_{n-[k/s]+1,n}}{%
X_{n-[k/s]+1,n}-X_{n-[k/s^{2}]+1,n}},
\end{equation*}
\begin{equation*}
\left(\frac{k}{n}\leq s^2 \leq 1\right),
\end{equation*}
which is  a generalization of the classical Pickands estimate $P_{n}(1/2)$ of the extremal index. We undertake here a purely stochastic process view for the 
asymptotic theory of that process by using the Cs\"{o}rg\H{o}-Cs\"{o}rg\H{o}-Horv\`{a}th-Mason (1986) \cite{cchm} weighted approximation of the empirical and quantile processes to suitable Brownian bridges. This leads to
the uniform convergence of the margins of this process to the extremal index and a complete theory of weak convergence of $P_n$ in $\ell^{\infty}([a,b])$ to some
Gaussian process 
$$
\left\{\mathbb{G},a\leq s \leq b\right\}
$$ 
for all $[a,b] \subset ]0,1[$. This frame greatly simplifies the former results and enable applications based on stochastic processes methods.
\end{abstract}

\maketitle

\section{Introduction}
\noindent Let $X_1,\ldots, X_n$ be a sequence of independant random variables of common distribution function $F$ with $F(1)=0$, and 
let $X_{1,n}<X_{2,n},\ldots<X_{n,n}$, denote the order statistics of $X_1,\ldots, X_n$ for any fixed $n\geq 1$. Let $k=k(n)$ be a sequence 
of positive integers satisfying:

\begin{equation}
 k \rightarrow +\infty, \quad  k/n\rightarrow 0, \quad and \quad \log \log n/k \rightarrow 0  \quad as\quad  n \rightarrow +\infty.  \tag{K}
\end{equation}

\noindent We are concerned with the following stochastic process
\begin{equation*}
P_{n}(s)=\log (1/s)^{-1}\log \frac{X_{n-k+1,n}-X_{n-[k/s]+1,n}}{%
X_{n-[k/s]+1,n}-X_{n-[k/s^{2}]+1,n}},\quad for\quad s^{2}\geq \frac{k}{n}.
\end{equation*}

\bigskip
\noindent If $s^{2}<\frac{k}{n}$, we put $P_{n}(s)=0$. For $s=1/2$, $P_{n}(1/2)$ is the so-called Pickands estimator of the extremal
index $\gamma$ when $F$ is in the extremal domain of a Generalized Extreme Value Distribution 
\begin{equation*}
 G_{\gamma}(x)=\exp(-(1 + \gamma x)^{-\frac{1}{\gamma}}),\quad for \quad 1 + \gamma x >0, \quad with \quad \gamma \in \mathbb{R}.
\end{equation*} 

\noindent This stochastic process has been studied by many authors: Alves (1995), Falk (1994), 
Pereira (1994), Yun (2000 and 2002), Segers (2002) \cite{segers}, etc. This latter gave a summarize of the previous works.

\bigskip
\noindent The handling of the Pickands process heavily depends that of the stochastic process of the large quantiles
\begin{equation}   
 \left\{X_{n-[k/s]+1,n}, n\geq 1\right\}. \label{proc1}
\end{equation}  
Drees (1995)\cite{dres1} and de Haan and Ferreira (2006)\cite{dehaan1} provided asymptotic uniform and Gaussian approximation of $(\ref{proc1})$ when $F$ is in the extremal domain under
second order conditions. Segers (2002) used such results to construct new estimators of the extremal index with integrals of statistics 
on the form of $P_{n}(s)$. This motivates us to undertake a purely stochastic process approach of the Pickands process in a simpler 
way but in a more adequate handling in order to derive from this study many potential applications. 

\section{Results}
\noindent Let us introduce some notation. First define the generalized
inverse of $F$:
$$F^{-1}(s)=\inf\left\{x, F(x) \geq s\right\}, 0 < s < 1,$$ and let
 \begin{equation*}
p_{n}(s)=\log (1/s)^{-1}\log \frac{F^{-1}(1-[k/s]/n)-F^{-1}(1-k/n)}{%
F^{-1}(1-[k/s^{2}]/n)-F^{-1}(1-[k/s]/n)}.
\end{equation*}
We are going to investigate this Pickands process
\begin{equation*}
 \left\{\kappa_n(s), \frac{k}{n}< s^2< 1 \right\}=\left\{\sqrt{k}( P_{n}(s)- p_{n}(s)), \frac{k}{n}< s^2< 1\right\}.
\end{equation*}

\noindent But it is easy to see that $P_n(s) \rightarrow K(\gamma)=\gamma\mathbf{1}_{\left\{\gamma \neq + \infty\right\}}$ for $F \in D(G_{1/\gamma})$. This
extends the motivation to the study of
\begin{equation*}
 \left\{\kappa^{*}_n(s), \frac{k}{n}< s^2< 1 \right\}=\left\{\sqrt{k}( P_{n}(s)- K(\gamma)), \frac{k}{n}< s^2< 1\right\}.
\end{equation*}

\noindent Since our conditions depend on auxilliary functions of the representations of functions $F$ in the extremal domain, that is $F\in D(G_{1/\gamma })$, $\gamma <0$, $\gamma >0$, $\gamma =+\infty$,
we feel obliged to introduce them. First for the Gumbel case $\gamma =+\infty ,$ we have
\begin{equation}
F^{-1}(1-u)=d-s(u)+\int_{u}^{1}t^{-1}s(t)dt  , \label{f1}
\end{equation}
where 
\begin{equation*}
s(u)=c(1+p(u))\exp (\int_{u}^{1}t^{-1}b(t)dt), 0< u < 1.
\end{equation*}
is a slowly varying function in the neighborhood of $0$. For the Frechet case $\gamma >0$, we have
\begin{equation}
F^{-1}(1-u)=c(1+p(u))u^{-K(\gamma) }\exp (\int_{u}^{1}t^{-1}b(t)dt). \label{f2}
\end{equation}

\noindent For the Weibull case $\gamma <0,$ it is known that $x_{0}(F)=\sup \{x,F(x)<1\}<\infty $ and we have 
\begin{equation}
x_{0}-F^{-1}(1-u)=c(1+p(u))u^{K(\gamma) }\exp (\int_{u}^{1}t^{-1}b(t)dt). \label{f3}
\end{equation}
In all these cases, $(p(u),b(u))\rightarrow (0,0)\text{ when }u\rightarrow 0$, $c$ is a positive constant  
\noindent The representations $(\ref{f2})$ and $(\ref{f3})$ are the Karamata representation, while $(\ref{f1})$ is that of de Haan\cite{dehaan}.

\bigskip
\noindent We fix two numbers $a$ and $b$, $a<b$, such that $[a,b]\subset ]0,1[$. For each $\gamma >0$, we will note $a(u)=F^{-1}(1-u)$ and
for $\gamma <0,$ $a(u)=x_{0}-F^{-1}(1-u)$. Set  
\begin{equation}
k_{a^{2}}=[a^{-2}]k
\end{equation}
and $\lambda >1$ a real number. Finally we define for an arbitrary function  $h$ defined on $(0,1)$ in $\mathbb{R},$%
\begin{equation*}
h_{n}(\lambda ,h,a)=\sup_{0\leq t\leq \lambda k_{a^{2}}/n}\left|
h(t)\right| .
\end{equation*}
We shall consider the regularity conditions 
\begin{equation}
\sqrt{k}p_{n}(\lambda ,p,a)\rightarrow 0, \quad as\quad  n \rightarrow +\infty.  \tag{RC1} \label{RC1}
\end{equation}
and
\begin{equation}
\sqrt{k}b_{n}(\lambda ,b,a)\rightarrow 0, \quad as\quad  n \rightarrow +\infty.  \tag{RC2} \label{RC2}
\end{equation}

\bigskip
\noindent All unspecified limits occur when $n\rightarrow
\infty.$ Finally we denote $o_{p}(s,a)$ et $o_{p}(s,a,b)$ respectively the uniform limits in $s\in \lbrack 0,1]$ and $s\in
\lbrack a,b].$ Here is our main results. 

\begin{theorem}\label{thm}
Let $F\in D(G_{1/\gamma })$,  $\gamma <0$, $\gamma >0$, $\gamma =+\infty$. Let $0<a<b<1$. If $(\ref{RC1})$ and $(\ref{RC2})$ hold
 then $\left\{\kappa_n(s), s \in [a,b]\right\}$ converges to a Gaussian process $\left\{\mathbb{G}(s), a< s< b \right\}$
in $\ell^{\infty}([a,b])$, of covariance function
\begin{align}
\Gamma(s,t)=& \frac{1}{(s^{-K(\gamma)}-1)(t^{-K(\gamma)}-1)\log s \log t}\{(s^{-K(\gamma)}-1)[t(t^{-K(\gamma)}-1) 
\nonumber \\
&+t^{-K(\gamma)}K(\gamma)s-K(\gamma)t^2 -K(\gamma)^2 t^2] + (t^{-K(\gamma)}-1)[K(\gamma)(s^{-K(\gamma)}t- s^2)] \nonumber \\
& + K(\gamma)^2 t^{-K(\gamma)}[s^{-K(\gamma)}-s^2]\},\nonumber 
\end{align}
with the convention that for $K(\gamma)=0$, $$\Gamma(s,t)=\frac{1-s^2}{(\log s)^2 (\log t)^2}.$$
Also $\left\{\kappa^{*}_n(s), a< s< b\right\}$ converges to $\mathbb{G}$, in $\ell^{\infty}([a,b])$.
\end{theorem}

\section{Proof of the results}
\noindent It is based on the so-called Hungarian construction of Cs\"{o}rg\H{o} et al. (1986) \cite{cchm}. For this define by $\{U_n(s),0 \leq s\leq 1$\}, the uniform empirical distribution function and $\{V_{n}(s)$, $0\leq s\leq 1\}$, the uniform empirical quantile function,  
based on the $n\geq 1$ first observations $U_1,U_2, \ldots$ sampled from a uniform random variables on $(0,1)$ and let $\left\{\beta_n(s); 0 \leq s \leq 1\right\}=\{\sqrt n (U_n(s)-s), 0 \leq s \leq 1 \}$ be the corresponding empirical process. One version of the Cs\"{o}rg\H{o} et al. (1986) \cite{cchm} is
the following 

\begin{theorem}[Cs\"{o}rg\H{o} et al. (1986)] \cite{cchm} \label{tcchm}
There exists a probability space holding a sequence of independant random variables $U_1, U_2, \ldots$ uniformly distributed on $(0,1)$ and a sequence
of Brownian bridges $$\left\{B_n(t), 0 \leq t \leq 1\right\}=\left\{W_n(t)-tW_n(1),  0 \leq t \leq 1 \right\},$$
where the $W_{n}$ are Wiener processes, such that for any $0<\nu <1/2$
 and $d \geq 0$
\begin{equation*}
\sup_{d/n\leq s\leq 1-d/n}\frac{| \beta _{n}(s)-B_{n}(s)| }{\left\{ s(1-s)\right\}
^{1/2-\nu }}=O_{p}(n^{-\nu }).
\end{equation*}
\end{theorem}
 
\bigskip
\noindent From this, we derive the Gaussian approximation of the following process
\begin{equation*}
\{U_{[k/s],n}-\frac{[k/s]}{n},\frac{k}{n}\leq s\leq 1\} 
\end{equation*}
in :

\begin{lemma} \label{lm1}
On the Cs\"{o}rg\H{o} et al. (1986) probability space we have
\begin{equation*} 
\sup_{k/(n-1)\leq s\leq 1}\left| \sqrt{k}(\frac{n}{[k/s]}%
U_{[k/s],n}-1)-W_{n}(1,s)\right| =O_{p}(1),
\end{equation*}
where $W_{n}(1,s)=s(\frac{k}{n})^{-1/2}W_{n}(k/ns)$ is a Wiener process.

\end{lemma}

\bigskip
\noindent Our proof is performed on the space of Theorem $\ref{tcchm}$. For conciseness, we restrict ourselves to the case $\gamma > 0$, since the other cases
are proved similary.

\bigskip
\noindent Let us begin by establishing that

\begin{equation} 
\frac{\sqrt{k}\left\{ X_{n-[k/s]+1,n}-F^{-1}(1-\frac{[k/s]}{n})\right\}}
{a(k/n)}=-K(\gamma) s^{K(\gamma) }W_{n}(1,s)+o_{p}(s,a,b). \label{f0} 
\end{equation}

\noindent Let
\begin{equation*}
A_{n}(s)=X_{n-k+1,n}-X_{n-[k/s]+1,n},\text{ }%
B_{n}(s)=X_{n-[k/s]+1,n}-X_{n-[k/s^{2}]+1,n}
\end{equation*}
\bigskip and 
\begin{equation*}
a_{n}(s)=F^{-1}(1-\frac{k}{n})-F^{-1}(1-\frac{[k/s]}{n}),\text{ }%
b_{n}(s)=F^{-1}(1-\frac{[k/s]}{n})-F^{-1}(1-\frac{[k/s^{2}]}{n}).
\end{equation*}

\bigskip
\noindent By using $(\ref{f2})$, we have
\begin{equation*}
\frac{X_{n-[k/s]+1,n}}{F^{-1}(1-\frac{[k/s]}{n})}=\frac{1+p(U_{[k/s],n})}{1+p([k/s]/n)}(\frac{n}{[k/s]}U_{[k/s],n})^{-K(\gamma)
}\exp (\int_{U_{[k/s],n}}^{[k/s]/n}t^{-1}b(t)dt).  \label{f01}
\end{equation*}
 
\bigskip
\noindent We shall treat the three items one by one.
Consider $s\in \lbrack a,1],$ $a>0.$ Then $%
[k/s]\leq \lbrack a^{-1}]k=k_{a}.$ Since $\frac{n}{k_{a}}%
U_{k_{a},n}\rightarrow 1$ in probability, then for each $\varepsilon
>0,$ for each $\lambda >1,$\ we have with probability at least greater than $%
1-\varepsilon ,$  for large values of n, 
\begin{equation*}
U_{k_{a},n}\leq \frac{\lambda k_{a}}{n}.
\end{equation*}
Since $(\ref{RC1})$ holds
\begin{equation*}
\sqrt{k}p_{n}(\lambda ,p,a)=\sqrt{k}\sup_{0<u<\lambda k_{a}/n}\left|
p(u)\right| \rightarrow 0,  
\end{equation*}
then
\begin{equation}
\sqrt{k}(\frac{1+p(U_{[k/s],n})}{1+p([k/s]/n)}-1)\rightarrow 0  \label{f03}
\end{equation}
in probability, uniformly in \ $s \in \lbrack a,b].$ Then, by using the Mean Value Theorem,
\begin{equation*}
(\frac{n}{[k/s]}U_{[k/s],n})^{-K(\gamma) }-1=-K(\gamma) (\frac{n}{[k/s]}%
U_{[k/s],n}-1)b_{n}(s),  \label{f04}
\end{equation*}
with 
\begin{equation}
b_{n}(s)\in \lbrack (\frac{n}{[k/s]}U_{[k/s],n})\wedge 1,(\frac{n}{[k/s]}%
U_{[k/s],n})\vee 1].  \label{bndes}
\end{equation}
Now by Lemma $\ref{lm1}$, uniformly in $s\in \lbrack a,b]$,%
\begin{equation*}
\sqrt{k}(\frac{n}{[k/s]}U_{[k/s],n}-1)=W_{n}(1,s)+O_{p}(1).
\label{f05}
\end{equation*}

\bigskip
\noindent Since $\sup_{s}\left| W_{n}(1,s)\right| $ is a finite random variable in probability, $\frac{n}{[k/s]}U_{[k/s],n}\rightarrow 1$
uniformly in probability and therefore $b_{n}(s)\rightarrow 1$
uniformly in probability. Let us set $o_{p}(s,a,b)$ as
meaning: tends to zero uniformly in $s\in \lbrack a,b]$. We have%
\begin{equation*}
(\frac{n}{[k/s]}U_{[k/s],n})^{-K(\gamma) }-1=-K(\gamma) (\frac{n}{[k/s]}%
U_{[k/s],n}-1)(1+o_{p}(s,a,b)).
\end{equation*}

\bigskip
\noindent It follows by applying Lemma $\ref{lm1}$ that, 
\begin{equation*}
\sqrt{k}(\frac{n}{[k/s]}U_{[k/s],n})^{-K(\gamma) }-1=-K(\gamma) W_{n}(1,s)+o_{p}(s,a,b).  \label{f06}
\end{equation*}

\bigskip
\noindent By the same techniques, we easily get for  $\varepsilon >0$, with probability at least greater than $1-\varepsilon$,
 
\begin{equation}
(\frac{n}{[k/s]}U_{[k/s],n})^{-b_{n}(\lambda ,a,b)}\leq \exp
(\int_{U_{[k/s],n}}^{[k/s]/n}t^{-1}b(t)dt)\leq (\frac{n}{[k/s]}%
U_{[k/s],n})^{b_{n}(\lambda ,a,b)}, \label{f07}
\end{equation}
for large values of $n$. We must show that both extreme terms tend to unity uniformly
in $s \in \lbrack a,b].$ Indeed
\begin{equation*}
(\frac{n}{[k/s]}U_{[k/s],n})^{b_{n}(\lambda ,a,b)}-1=b_{n}(\lambda ,a,b)(%
\frac{n}{[k/s]}U_{[k/s],n}-1)b_{n}(s),
\end{equation*}
where $b_{n}(s)$ is already defined in (\ref{bndes}) and is $%
o_{p}(s,a,b).$ We arrive at
\begin{equation*}
\sqrt{k}\left\{ (\frac{n}{[k/s]}U_{[k/s],n})^{b_{n}(\lambda ,a,b)}-1\right\}=b_{n}(\lambda ,a,b)(W_{n}(1,s)+o_{p}(s,a,b))(1+o_{p}(s,a,b))
\end{equation*}
 
\bigskip
\noindent which is an $o_{p}(s,a,b)$ term, since $\sup_{s\in (a,b)}\left|
W_{n}(1,s)\right| $ is bounded in probability. Further, put 

\bigskip
\noindent $A_{1,n}(s)=\frac{1+p(U_{[k/s],n})}{1+p([k/s]/n)}$, $A_{2,n}(s)=(\frac{n}{[k/s]}U_{[k/s],n})^{-K(\gamma) }$ and $A_{3,n}(s)=\exp (\int_{U_{[k/s],n}}^{[k/s]/n}t^{-1}b(t)dt)$. 

\bigskip
\noindent Now we have
\begin{equation*}
\frac{X_{n-[k/s]+1,n}}{F^{-1}(1-\frac{[k/s]}{n})}-1 =A_{2,n}(s)A_{3,n}(s)\left\{ A_{1,n}-1\right\} +A_{2,n}\left\{
A_{3,n}-1\right\} +\left\{ A_{2,n}-1\right\} . \label{f08}
\end{equation*}
This leads to
\begin{align}
&\sqrt{k}A_{2,n}(s)A_{3,n}(s)\left\{ A_{1,n}-1\right\} +A_{2,n}\left\{
A_{3,n}-1\right\} +\left\{ A_{2,n}-1\right\} \nonumber \\
&=A_{2,n}(s)A_{3,n}(s)\sqrt{k}\left\{ A_{1,n}-1\right\} +A_{2,n}\sqrt{k}%
\left\{ A_{3,n}-1\right\} +\sqrt{k}\left\{ A_{2,n}-1\right\}. \nonumber
\end{align}

\bigskip
\noindent Finally, we arrive at
\begin{equation*}
\frac{\sqrt{k}\left\{ X_{n-[k/s]+1,n}-F^{-1}(1-\frac{[k/s]}{n})\right\} }{%
a([k/s]/n)} =-K(\gamma) W_{n}(1,s)+o_{p}(s,a,b), 
\end{equation*}
 
\bigskip
\noindent where, for $a(s)=F^{-1}(1-s)$, we used the result that $a([k/s]/n)/a(k/n)$
tends uniformly in $s^{-K(\gamma) }$ for $s \in (a,b).$ This achieves the proof of $(\ref{f0})$.

\bigskip
\noindent Now we use $(\ref{f0})$ to prove Theorem $\ref{thm}$. Put

\bigskip
$W_{n}(2,s)=(\frac{k}{n})^{-1/2}W_{n}(k/n)$, $W_{n}(3,s)=s^{2}(\frac{k}{n})^{-1/2}W_{n}(s^{-2}k/n).$
 
\bigskip
\noindent This latter is a Gaussian process with covariance function $\min (s^{2},t^{2}).$ Denote now
\begin{equation*}
C_{n}(s)=\frac{A_{n}(s)}{B_{n}(s)}=\frac{X_{n-[k/s]+1,n}-X_{n-k+1,n}}{%
X_{n-[k/s^{2}]+1,n}-X_{n-[k/s]+1,n}}
\end{equation*}
and 
\begin{equation*}
c_{n}(s)=\frac{a_{n}(s)}{b_{n}(s)}=\frac{F^{-1}(1-\frac{[k/s]}{n})-F^{-1}(1-\frac{k}{n})}{%
F^{-1}(1-\frac{[k/s^{2}]}{n})-F^{-1}(1-\frac{[k/s]}{n})}.
\end{equation*}
Then
\begin{equation*}
\log C_{n}(s)-\log c_{n}(s)=(C_{n}(s)-c_{n}(s))\times c_{n}(1,s)^{-1},
\label{d01}
\end{equation*}
where 
\begin{equation*}
c_{n}(1,s)\in \lbrack C_{n}(s)\wedge c_{n}(s),C_{n}(s)\vee
c_{n}(s)]\rightarrow s^{-K(\gamma) },
\end{equation*}
uniformly in $s\in (0,1)$ and $c_{n}(1,s)^{-1}\rightarrow s^{K(\gamma) }$
uniformly in $s\in \lbrack a,1].$ Then 
\begin{equation*}
\log C_{n}(s)-\log c_{n}(s)=(s^{K(\gamma) }+o_{p}(s,a))\text{ }%
(C_{n}(s)-c_{n}(s)).
\end{equation*}
Next 
\begin{equation*}
C_{n}(s)-c_{n}(s)=\frac{%
-a_{n}(s)(B_{n}(s)-b_{n}(s))+b_{n}(s)(A_{n}(s)-a_{n}(s))}{b_{n}(s)B_{n}(s)}.
\label{d02}
\end{equation*}

\bigskip
\noindent The same techniques leads to
\begin{equation*}
a_{n}(s)=a([k/s]/n))(1-s^{-K(\gamma) })(1+o_{p}(s,a)),\text{}%
A_{n}(s)=a(U_{[k/s],n})(1-s^{-K(\gamma) })(1+o_{p}(s,a))
\end{equation*}
and 
\begin{equation*}
b_{n}(s)=a([k/s^{2}]/n)(1-s^{-K(\gamma) })(1+o_{p}(s,a)),\text{}%
B_{n}(s)=a(U_{[k/s^{2}],n})(1-s^{-K(\gamma) })(1+o_{p}(s,a)).
\end{equation*}
Next 
\begin{equation*}
B_{n}(s)-b_{n}(s) =a([k/s^{2}]/n)(\frac{X_{n-[k/s^{2}]+1,n}}{F^{-1}(1-[k/s^{2}]/n)}%
-1)-a([k/s]/n)(\frac{X_{n-[k/s]+1,n}}{F^{-1}(1-[k/s]/n)}-1).
\end{equation*}
When gathered, these formulas imply
\begin{align}
\sqrt{k}(B_{n}(s)-b_{n}(s))&=a([k/s^{2}]/n)(1+o_{p}(s,a))(-K(\gamma)
W_{n}(3,s)+o_{p}(s,a)) \nonumber \\
&-a([k/s]/n)(1+o_{p}(s,a))(-K(\gamma) W_{n}(1,s)+o_{p}(s,a)) \nonumber
\end{align}
and 
\begin{equation*}
\sqrt{k}\frac{-a_{n}(s)(B_{n}(s)-b_{n}(s))}{b_{n}(s)B_{n}(s)} 
=-K(\gamma) \frac{s^{-K(\gamma) }}{(1-s^{-K(\gamma) })}W_{n}(3,s)+K(\gamma) \frac{%
s^{-2K(\gamma) }}{(1-s^{-K(\gamma) })}W_{n}(1,s)+o_{p}(s,a).
\end{equation*}

\noindent Similarly, we have
\begin{equation*}
\sqrt{k}\frac{b_{n}(s)(A_{n}(s)-a_{n}(s))}{b_{n}(s)B_{n}(s)}
=-K(\gamma) \frac{s^{-K(\gamma) }}{(1-s^{-K(\gamma) })}W_{n}(1,s)+K(\gamma) \frac{%
s^{-2K(\gamma) }}{(1-s^{-K(\gamma) })}W_{n}(2,s)+o_{p}(s,a).
\end{equation*}
 
\bigskip
\noindent We conclude that 
\begin{equation*}
(\log (1/s))^{-1}\sqrt{k}\left\{ \log C_{n}(s)-\log c_{n}(s)\right\}=\log (1/s)^{-1}\frac{K(\gamma)}{(1-s^{-K(\gamma) })}%
(1+o_{p}(s,a))\times
\end{equation*}
\begin{equation*}
\left\{ (s^{-K(\gamma) }-1)(W_{n}(1,s)+o_{p}(s,a))+s^{-K(\gamma)
}(W_{n}(2,s)+o_{p}(s,a))-W_{n}(3,s)+o_{p}(s,a)\right\} .
\end{equation*}

\bigskip
\noindent Now by restraining ourselves to $s\in \lbrack a,b]\subset ]0,1[,$ we get uniformly in
those $s,$

\begin{equation*}
\kappa_n(s)=\sqrt{k}\left\{ P_{n}(s)-\log c_{n}(s)/\log (1/s)\right\}
=\mathbb{G}_{n}(s)+o_{p}(s,a,b),
\end{equation*}
where 
\begin{equation*}
\mathbb{G}_{n}(s)=\frac{K(\gamma) }{(s^{-K(\gamma) }-1)\log s}\left\{ (s^{-K(\gamma)
}-1)(W_{n}(1,s)+s^{-K(\gamma) }W_{n}(2,s)-W_{n}(3,s)\right\}
\end{equation*}
is a Gaussian process with covariance $\Gamma_n(s,t)\rightarrow \Gamma(s,t)$ expressed in Theorem $\ref{thm}$. To extend our result to $\kappa^*_n(s)$ we have to prove 
that under $(\ref{RC1})$ and $(\ref{RC2})$, \begin{equation}\sqrt{k}(p_n(s)-K(\gamma))\rightarrow 0 \label{eq}\end{equation}
In the Theorem $\ref{thm}$, the asymptotic laws comes from that of $\frac{n}{k}U_{[k/s],n}$. All the remainder terms are controlled uniformly by the regularity conditions. If we treat $(\ref{eq})$
the corresponding part $\frac{n}{k}[k/s]/n$ satisfies : $$\sqrt{k}(\frac{n}{k}[k/s]/n - (\frac{1}{s})^{K(\gamma)})\rightarrow 0$$
uniformly in $s \in [a,b]$. By respectively the same proofs, we arrive at  the result for $\gamma < 0$, $\gamma>0$, $\gamma=+\infty$,
and this completes the proof.

\section{Applications}
\noindent Consider a measure $m$ bounded on compacts set $[a,b] \subset ]0,1[$ and define as Segers(2002) \cite{segers} the statistics, with $K(\gamma)=\gamma1_{\gamma\neq + \infty}$, 
$$\mathbb{I}_n(a,b,m)=\int_{a}^{b}P_n(s)dm(s)$$
We get 
\begin{align}\sqrt{k}(\mathbb{I}_n(a,b,m)-K(\gamma))&=\int_{a}^{b}\sqrt{k}(P_n(s)-K(\gamma))dm(s)\nonumber \\
&=\int_{a}^{b}\mathbb{G}_n(s)dm(s) \rightarrow \int_{a}^{b}\mathbb{G}(s)dm(s)=\mathbb{I}(a,b,m) \nonumber \end{align}
\begin{theorem}
 Let $0<a<b<1$, and $m$ a measure bounded on compact set $[a,b] \subset ]0,1[$. Let $F \in D(G_{1/\gamma})$, $\gamma < 0$, $\gamma>0$, $\gamma=+\infty$, and 
$(\ref{RC1})$ and $(\ref{RC2})$ holds. Then 
$$\sqrt{k}(\mathbb{I}_n(a,b,m)-K(\gamma))\rightarrow \mathcal{N}(0, \sigma_m^2)$$
with 
$$\sigma_m^2=\mathbb{E}\int_{a}^{b}\mathbb{G}(s)dm(s)=\int_{a}^{b}\int_{a}^{b}\Gamma(s,t)dm(s)dm(t)$$
\end{theorem}
\noindent We can investigate the problem of minimizing $\sigma_m^2$, that is finding
$$\min_{m \in \mathcal{M}}\int_{a}^{b}\int_{a}^{b}\Gamma(s,t)dm(s)dm(t)$$
in future works. Remind that for $m_s=\delta_s$, we have $\mathbb{I}_n(a,b,m_s)=P_n(s)$.

\section{Continuity modulus}
\begin{lemma}
 We have, almost surely, for any $s \in [a,b]\subset ]0,1[$
\begin{equation*}
 \lim_{h\rightarrow 0}\sup_{|s-t|\leq h}\frac{|\kappa_n(s)-\kappa_n(t)|}{w(h)}\leq L(\gamma)
\end{equation*}
with 
\begin{equation*}
L(\gamma)= \frac{K(\gamma)}{|\log b|} + \frac{K(\gamma)}{(|b^{-K(\gamma)}-1||\log b|)},  \label{lm2}
\end{equation*}
with the convention that $$L(\gamma)=\frac{1}{(|\log b|)^2} \quad for \quad K(\gamma)=0$$
and
$w(h)=\sqrt{2h\log(1/h)}$
\end{lemma}
\noindent In other words, the Gaussian process $\kappa_n(s)$ have modulus of continuity $w_{\kappa}(\sqrt{2\delta\log(1/\delta)})$ with probability $1$ and for
sufficiently small $\delta > 0$.

\begin{proof}
 Put
$$g(s)=\frac{K(\gamma) }{(s^{-K(\gamma)}-1)\log s}$$ and
$$h_n(s)=(s^{-K(\gamma) }-1)W_{n}(1,s)+s^{-K(\gamma)}W_{n}(2,s)-W_{n}(3,s).$$

\bigskip
\noindent We have the following decomposition
\begin{equation*}
 |\mathbb{G}_n(s)-\mathbb{G}_n(t)|=|(g(s)-g(t))h_n(s)| + |g(t)(h_n(s)-h_n(t))|
\end{equation*}
Now by using the exact continuity modulus of the Wiener Process $W_n(.,s)$, and by repeatingly using the Mean Value Theorem, we easily arrive at the results. We express that these results hold for a compact set $[a,b], 0< a< b<1$.
\end{proof}

\noindent \textbf{Acknowledgement}.
The paper was finalized while the second author was visiting MAPMO, University of Orl\'eans, France, in 2011. She expresses her warm thanks to responsibles of MAPMO for kind hospitality. She is also granted by the \textit{Coop\'eration Fran\c{c}aise} (EGIDE) and by \textit{AIRES-Sud}, a programme from the French Ministry of Foreign and European Affairs implemented by the \textit{Institut de Recherche pour le D\'eveloppement (IRD-DSF)}\\

\end{document}